\documentclass[aps, pra, 10pt, twoside, twocolumn, superscriptaddress, showkeys, floatfix]{revtex4-1}
\pdfoutput=1
\usepackage{wrapfig}
\usepackage{upgreek}
\usepackage{latexsym,amssymb}
\usepackage[intlimits]{amsmath}
\usepackage{graphicx}
\usepackage{float}
\usepackage{soul}
\usepackage[normalem]{ulem}
\usepackage{color}
\usepackage{natbib}
\usepackage{hyperref}
\hypersetup{
     colorlinks   = true,
     citecolor    = blue
}
\def\be{\begin{equation}}
\def\ee{\end{equation}}
\def\e#1{\label{#1}\end{equation}}
\def\bea{\begin{eqnarray}}
\def\eea{\end{eqnarray}}
\def\ea#1{\label{#1}\end{eqnarray}}

\def\bem#1{\begin{mathletters}\label{#1}}
\def\eml{\end{mathletters}}

\def\ket#1{{|#1\rangle}}
\def\bra#1{{\langle#1|}}

\def\4#1{{\boldsymbol{#1}}}
\def\8#1{{\widetilde{#1}}}
\def\bse{\begin{subequations}}
\def\ese{\end{subequations}}

\def\ketbra#1#2{\ket{#1}\bra{#2}}

\def\0{\ket{0}}
\def\1{\ket{1}}

\begin{document}
\title{Quantum sensing and control of spin state dynamics in the radical pair mechanism}
\author{Amit Finkler}
\affiliation{Department of Chemical and Biological Physics, Weizmann Institute of Science, Rehovot 7610001, Israel}
%\orcid{0000-0002-9932-8270}
\author{Durga Dasari}
\affiliation{3.\,Physikalisches Institut, Universit\"at Stuttgart, 70569 Stuttgart, Germany}
%\orcid{0000-0002-7899-6755}
\date{July 31, 2020}

\begin{abstract}
Radical pairs and the dynamics they undergo are prevalent in many chemical and biological systems. Specifically, it has been proposed that the radical pair mechanism results from a relatively strong hyperfine interaction with its intrinsic nuclear spin environment. While the existence of this mechanism is undisputed, the nanoscale details remain to be experimentally shown. We analyze here the role of a quantum sensor in detecting the spin dynamics (non-Markovian) of individual radical pairs in the presence of a weak magnetic field. We show how quantum control methods can be used to set apart the dynamics of radical pair mechanism at various stages of the evolution. We expect these findings to have implications to the understanding of the physical mechanism in magnetoreception and other bio-chemical processes with a microscopic detail.

\end{abstract}

\maketitle

\section{Introduction}
Spin plays a fundamental role in many chemical reactions, from photosynthesis \cite{Norris1971} to polymerization \cite{Yamada1999}. One of the most well-studied of those is known as the radical-pair mechanism \cite{Steiner1989}. There, two radicals, i.e., molecules with a free electron ($S=1/2$), are brought by excitation to such a close distance between them, that their two electrons become entangled and emerge, depending on minimum energy considerations, as either in a triplet or in a singlet \cite{Kaptein1969} (in both cases, $S=1$). Due to interaction with an external magnetic field and the existence of hyperfine interaction with each radical's nuclear spins, this pair of radicals can undergo oscillations between its triplet and singlet states, with a frequency dependent mostly on the strength of this hyperfine coupling with respect to the surrounding magnetic field \cite{Anisimov1983}. 
The radical-pair mechanism had been studied extensively using recombination fluorescence \cite{Stass2000}, electrochemistry \cite{Zeng2018}, transient ESR \cite{Biskup2009, Weber2010a} and ultrafast absorption spectroscopy \cite{Liedvogel2007, Maeda2008, Kerpal2019}. All of these techniques probe samples of macroscopic scale, namely microliters or $10^{18}$ molecules. While providing a wealth of information and improving our understanding of some key processes in said mechanism, the aforementioned tools provide an ensemble average indication as to the magnetic properties one wants to explore. The radical pair mechanism is typically characterized by the spin coherence time \cite{Kominis2009, Cai2010, Gauger2011, Kattnig2016} and the recombination rate~$\tilde{\kappa}$, i.e., the rate at which the pairs recombine back to their original constituents \cite{Rodgers2009}.
Our approach, using a nanoscale single-spin sensor in the form of the nitrogen vacancy center in diamond \cite{Gruber1997, Taylor2008, Cai2013b}, shows that some of the most remarkable features of the RPM are masked by averaging, and allows us to probe this as-of-now yet un-chartered territory. 
Our quantum-sensing technique can assist us in determining not only the charge state of the pair \cite{Liu2017a} but perhaps more importantly, its spin state. We introduce a detection pulse scheme for single qubit magnetometry \cite{Balasubramanian2008, Maze2008, Taylor2008}, that enables a consistent way of realizing whether the pair is in its singlet or one of its three possible triplet states.
As this is a single-spin sensor operated in a detection regime where only a small amount of molecules (and hence radical pairs or spins) contributes to the signal \cite{Grotz2011, Grinolds2014, Schlipf2017}, we also show how quantum control schemes allow us to modify and in some instances also enhance the interaction of the radical pair with its external environment, thereby achieving a change in the ratio between the final products.
\begin{figure*}[ht!]
    \centering
    \includegraphics[width = 1.0\textwidth]{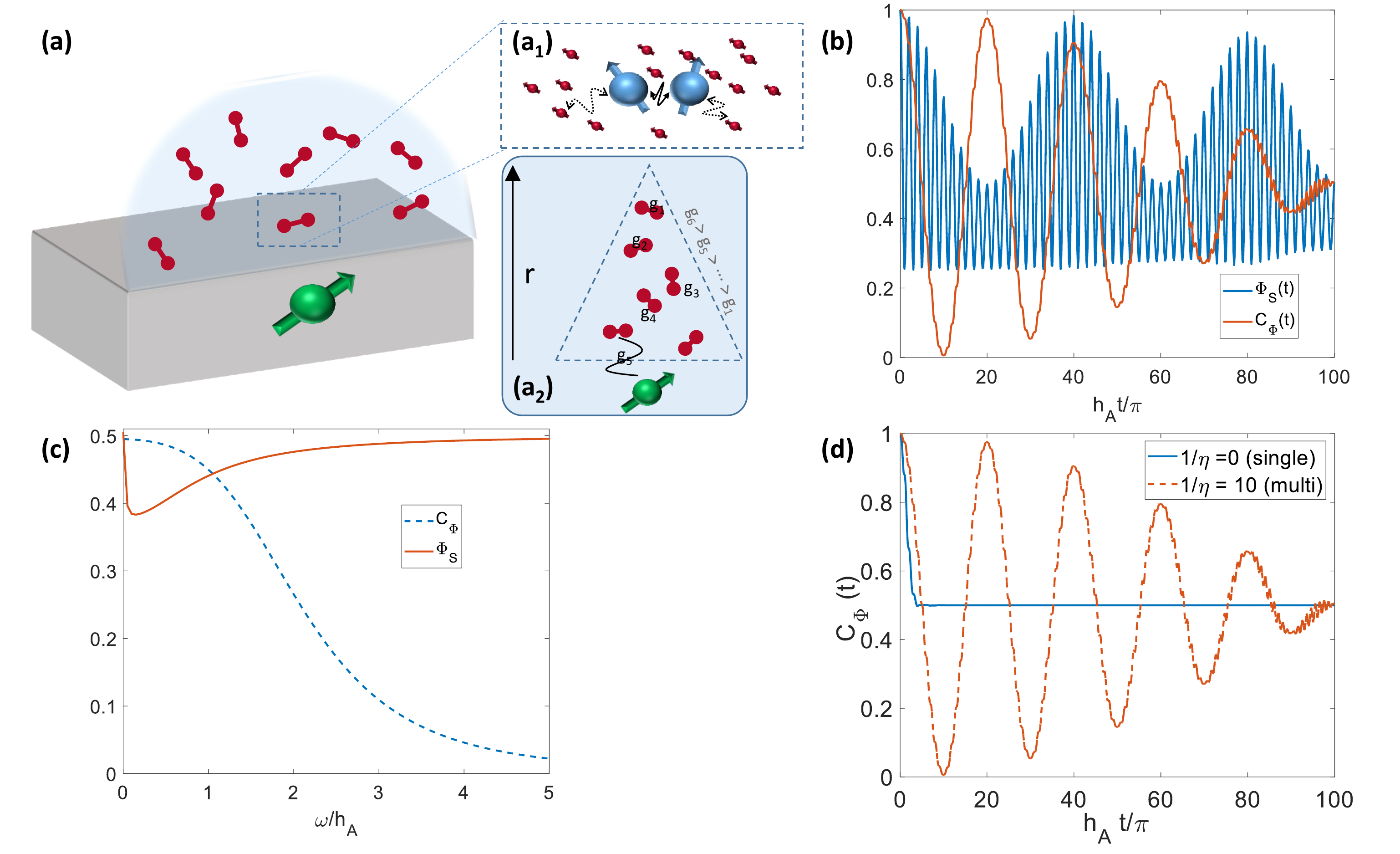}
    \caption{\textbf{a.} Schematic representation of the system, wherein the radical pair spins deposited on the surface of the diamond are sensed/controlled by a single spin sensor embedded in a solid state matrix. The inset ($a_1$) shows the coupling of radical pair spins to their local nuclear spin environment and in ($a_2$) the distance-dependent coupling of NV to  the radical pair spins. \textbf{b.} The singlet-triplet oscillations and the corresponding oscillations in the contrast of the sensor are shown as a function of time. \textbf{c.} The asymptotic singlet state population and the sensor spin contrast  (Eq.~\eqref{eq:fractions}) are plotted as a function of a weak external field $\omega$. \textbf{d.} The average singlet production yield in the presence of the sensor-induced gradient field is shown for the cases of zero width ($\eta$) and finite width of the distribution of the coupling strengths $g$ (see Eq.~\eqref{eq:contrast_eta}). 
    In the above simulations we have chosen  all the parameters in units of the hyperfine coupling strength $h_A$: the singlet recombination rate $\tilde{\kappa} = 0.01 h_A$, and the sensor-RP coupling to be $g = 0.1 h_A$.  }
    \label{fig:theoretical}
\end{figure*}

\section{Model}
We consider a prototypical model used to analyze the sensitivity of a radical pair reaction to external field  ($\omega$), wherein a radical pair ($\sigma$), composed of two electrons (spin-$1/2$) coupled to a nuclear spin environment ($I$). In addition, we also couple the radical pair spins to a sensor (spin-$1$), $S$. The Hamiltonian describing their dynamics is given by
\begin{equation}
H = h_A\vec{I_A}\cdot{\sigma_A} + h_B\vec{I_B}\cdot{\sigma_B} + \omega (\sigma^z_A + \sigma^z_B) + g S^z (\sigma^z_A + \sigma^z_B),
\label{eq:hamiltonian}
\end{equation}
where $h_{A,B}$ are the hyperfine coupling constants and $g$ is the RP-sensor coupling strength. For simplicity we set $h_B=0$, and consider a single nuclear spin as the dynamics generated by the above Hamiltonian becomes exactly solvable \cite{Timmel1998, Hogben2012}.  One could choose a two-level subspace spanned either by $\ket{-1}, \ket{0}$ or $\ket{+1}, \ket{0}$ as our computational subspace for the sensor spin.
For example, in this basis of the sensor $\ket{\pm1}, \ket{0}$, the above Hamiltonian can be further simplified to 
\be
H = H_0\ket{0}\bra{0} + H_\pm\ketbra{\pm 1}{\pm 1}.
\label{eq:simp_H}
\ee
Here $H_0 = h_A\vec{I}\cdot{\sigma_A} + \omega (\sigma^z_A + \sigma^z_B)$, and $H_\pm = h_A\vec{I}\cdot{\sigma_A} + (\omega \pm g) (\sigma^z_A + \sigma^z_B)$. From the above equation it is clear that the coupling to the sensor leads to a modified external field for the radical pair. Due to the sensor spin-state dependent enhancement (reduction) of the effective external field seen by the radical pair, the singlet product yield $\Phi_S$ varies with the occupation probability of the sensor in either of its spin states. In turn, this leads to a visible contrast in the spin-state readout of the sensor itself. Following Ref.~\cite{Timmel1998} and evaluating the singlet fraction, $\Phi^0_S, \Phi^\pm_S$ for the corresponding sensor spin-states, we find the modified singlet fraction yield, $\Phi_S$ 
\be
\Phi_S = (\Phi^\pm_S + \Phi^0_S)/2, ~~ C_\Phi \approx |\Phi^\pm_S - \Phi^0_S|.
\label{eq:fractions}
\ee
Here $\Phi_S$ is obtained by tracing out sensor degrees of freedom from exact dynamics using the above Hamiltonian. The sensor signal (contrast $C_\Phi$) is obtained by a Ramsey measurement, i.e., starting from the sensor state which is in a equal quantum superposition of its spin states ($\pi/2$-pulse), and letting it evolve freely under the evolution generated by Eq.\eqref{eq:simp_H}. After the free-evolution another $\pi/2$-pulse is applied to map the phase accumulated during the free evolution to the population of the spin-states which is then measured to obtain $C_\Phi$ as the population difference between the two states.
After tracing out the RP degrees of freedom (see Appendix for details), the time-dependent sensor contrast is given by
\be
C_\Phi = \mathrm{Tr}\left[Re\left\{{\rm e}^{iH_0 t} \rho_\mathrm{RP}(0)\otimes\rho_I(0){\rm e}^{iH_1 t}\right\}\right]
\ee
where $\rho_\mathrm{RP}(0) = \ket{S}\bra{S}$ is the initial singlet state of the radical pair spins, and $\rho_I(0) = 1/2 \bf{I}$ is the initial thermal state of the nuclear spin. One can obtain an exact expression upon simplifying the above equation, given by

\begin{widetext}
\begin{multline}
C_\Phi(t) = \frac{1}{\Omega_1\Omega_2}\left[\sin\left(t\Omega_1\right)\left(2\left(\Omega_1^2+g\omega\right)\cos\left(gt\right)\sin\left(t\Omega_2\right)-2\omega\Omega_2\sin\left(gt\right)\cos\left(t\Omega_2\right)\right)\right. \\
+\left.\Omega_1\cos\left(t\Omega_1\right)\left(2\left(g+\omega\right)\sin\left(gt\right)\sin\left(t\Omega_2\right)+2\Omega_2\cos\left(gt\right)\cos\left(t\Omega_2\right)\right)+2\Omega_1\Omega_2\right]
\end{multline}

\end{widetext}
where $\Omega_1 = \sqrt{h_A^2+\omega^2}$ and $\Omega_2 = \sqrt{h_A^2+(\omega+g)^2}$.
This is the central result of the paper and from this we obtain directly the dynamic behavior shown in Fig.~\ref{fig:theoretical}(b). The singlet recombination rate ${\kappa}$ determines how fast the RP state is reset back to the singlet state before a significant hyperfine-driven singlet to triplet conversion can take place. For finite ${\kappa}$, the sensor contrast, including its own relaxation rate $\gamma$, can be simply approximated as 

\be
C_\Phi(\tilde{\kappa}) \approx \int^\infty_0 \mathrm{d}t C_\phi(t){\rm e}^{-\tilde{\kappa} t}, 
\ee

where $\tilde{\kappa} = {\kappa} + \gamma$. The above integral is exactly solvable (see Appendix) and can be further used for sensitivity analysis of the sensor with respect to the changes both in the external field ($\omega$) and the coupling strength~($g$) to the RP. As the coherence time of the sensor is much longer than the singlet recombination rate ${\kappa}$, the asymptotic dependence of both the singlet fraction and the sensor contrast ($C_\phi$) on the external field $\omega$ as shown in Fig.~\ref{fig:theoretical}(c) will be dominantly determined by ${\kappa}$ alone.
In Fig.~\ref{fig:theoretical}(c) we show the effect of magnetic field $(\omega)$ on the singlet yield of a one-proton radical pair (see Eq.~\eqref{eq:hamiltonian}). It has been shown previously that the singlet product yield shows an abrupt change even by a tiny magnetic field in the low-field limit and for a slow recombination rate~${\kappa}$~\cite{ Kattnig2016, Maeda2012}. We observe a similar behavior in the sensor contrast, $C_\phi$, with $\omega$. The singlet fraction shown here is the steady state population obtained in the long-time limit where the coherent singlet-triplet mixing and the incoherent mixing caused by the relaxation are in equilibrium. One could also analyze the dynamics for times shorter than the relaxation time as shown in  Fig.~\ref{fig:theoretical}(c). The singlet-triplet oscillations are also seen on the sensor contrast.  As opposed to the fast oscillations in the singlet fraction caused by the hyperfine interaction $h_A$, the oscillation frequency of the sensor contrast is due to the small coupling strength  $g \ll h_A$.
\begin{figure*}[ht!]
    \includegraphics[width = 1.0\textwidth]{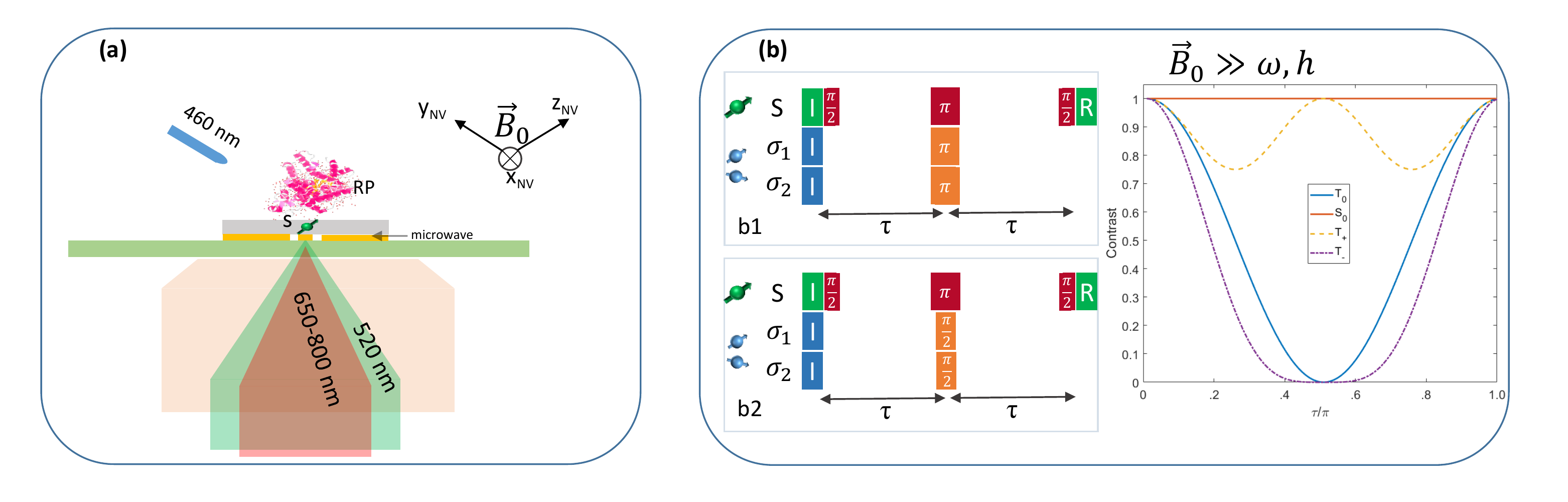}
    \caption{ \textbf{a.}\ A schematic representation of the experimental setup displaying a diamond substrate with RP spins on its surface. We also show the initialization and readout mechanism of the sensor spin (NV) state through the green laser ($520$ nm) and detection through its red side-band emission ($650-800$ nm). A blue LED ($460$ nm) is used to initialize the RP mechanism that will be sensed through the NV center in the diamond lattice. \textbf{b.}\ We show the modified DEER pulse sequence ($b_1$) in $b_2$ wherein a $\pi$-pulse is replaced by a $\pi/2$-pulse to differentiate the singlet and triplet states as shown in the plot on the right, where we vary the free evolution time $\tau$, and initialize the RP in various spin states.}
    \label{fig:experimental}
\end{figure*}
\section{Ensemble sensing}
While the above analysis holds for a single RP with a given interaction strength $g$ to the sensor holds in general, in a practical setting, an ensemble of RPs is dropcasted on top of the diamond surface or on the apex of an AFM tip and then scanned with respect to the sensor \cite{Schmid-Lorch2015}. Due to this the sensor interacts with a large number of RPs within a sensing volume that is determined by the distance between the surface and the sensor \cite{Staudacher2013}. Owing to interaction with a large number of RPs both the singlet production rate and the sensor contrast get affected due to the averaging over the effective field generated by the sensor on the different RPs and vice-versa. Such effective fields produced by a RP spin-bath could be approximated in the quasi-static approximation by a Gaussian distribution \cite{Prokof2000} when the effective field produced by the bath is varying slowly in time. Upon using such a distribution for $g$ one can integrate the functions given in Eq.~\eqref{eq:fractions}, as 
\begin{align}
    \Phi_S(\eta) & = \int \mathrm{d}g {\rm e}^{-\eta(g-g_0)^2}\Phi_S(g),\nonumber \\ C_\Phi(\eta) & = \int \mathrm{d}g {\rm e}^{-\eta(g-g_0)^2}C_\Phi(g).
    \label{eq:contrast_eta}
\end{align}
For increasing width of this distribution the oscillations seen in the single RP limit eventually vanishes in the large $N$-limit. A similar behavior can also be seen for finite time in Fig.~\ref{fig:theoretical}(c), where we directly plot the time-dependent behavior of the triplet-pair production $\Phi(t)$ from the unitary evolution generated by Hamiltonian given in Eq.~\eqref{eq:hamiltonian}. The characteristic oscillations of the triplet fraction shown in the Fig.~\ref{fig:theoretical}(d) indicate the singlet-triplet oscillations. 
In the presence of interaction with the sensor, the dampening of the oscillations can be understood as an additional random phase $\phi$ (see Appendix) introduced by the sensor coupling, leading to a gradual loss of coherent behavior of the triplet-pair production in time. Both the asymptotic and finite time analysis clearly display the role of an additional sensor interaction both for sensing the RP dynamics, and in turn influencing its production rate. This influence can be further induced in a controlled manner if one employs coherent control of the sensor spin (see Appendix). 

\section{Experimental Implementation}
Apart from the theoretical model shown here, we also propose an apparatus for sensing and controlling the RP spin state as described above. This consists of a scanning confocal microscope for the initialization and read-out of a single nitrogen-vacancy (NV) center in diamond \cite{Gruber1997, Jelezko2004a, Balasubramanian2008}, as is schematically shown in Fig.~\ref{fig:experimental}(a). Green laser (520 nm) is used for excitation, and the NV fluorescence in the range of 650 nm to 800 nm is collected and focused onto an avalanche photo-diode. The microscope is integrated with a variable magnetic field implemented by either a permanent magnet (NdFeB or SmCo) on an XYZ stage for a room temperature setup or a set of three pairs of split Helmholtz coils for a low temperature measurement. In addition, a microwave antenna is positioned near the NV center for spin state manipulation of both the NV spin and the RP state. A schematic of such an apparatus is shown in Fig.~\ref{fig:experimental}(a). A solution containing a dilute amount of RPs can be drop-casted on the surface of a diamond containing shallow NVs. For such sensors, situated approximately 5 to 10 nm below the surface of the diamond \cite{Favaro2017}, the typical decoherence time (spin echo) is $T_2 \sim 10-20\ \upmu\mathrm{s}$. Finally, we take RPs based on the Flavin-Tryptohpan pair in cryptochrome as a widely-studied example\cite{Biskup2009, Bialas2016}, where a pulsed blue diode (460 nm) can be used to excite RPs. This wavelength was shown to be efficient in creating RPs in this makeup. Not only that, but cryptochrome's absorption spectra is coincidentally ``orthogonal'' to the NV's \cite{Song2006, Biskup2009}, preventing optical signal quenching which might reduce the SNR. Spin relaxation times for this RP system are estimated to be in the range of a few $\upmu\mathrm{s}$, consistent with transient absorption experiments \cite{Maeda2012}. For a realistic NV ac magnetic field sensitivity of 10 nT/Hz$^{1/2}$ and a distance of 20 nm between the NV and the RP, we expect to be able to achieve an SNR of 1 after a measurement time of 10 seconds \cite{Taylor2008, Grinolds2013}. As each iteration of our pulse sequence is on the order of 10 $\upmu\mathrm{s}$, and with a single-shot SNR 0.03 (for a single readout pulse), each data point in the scheme proposed above (see Fig.\,\ref{fig:experimental}b) needs at least $10^5$ repetitions, or 1 second, to achieve an SNR of 10 \cite{Hopper2018}. When taking extra measures for canceling sources of noise \cite{HaeberleT.2015} and considering the reduced contrast (and hence, SNR) in nested NV magnetometry measurements \cite{Staudacher2013, Glenn2018}, a full data set would be acquired in about one hour. See Appendix for a more detailed calculation.

\section{Quantum control}
\color{black} 
As both the sensor and the target spins are electrons, one can simultaneously manipulate the sensor and target spins similar to the double electron-electron resonance experiments known in NMR and EPR. In contrast here as we have two target electron spins, one can employ a triple-electron resonance (TEER) sequence as an alternative method to detect the spin state of the radical pair. We have introduced and experimentally verified such multi-electron resonance control earlier in Ref.~\cite{Schlipf2017}. This entails a preliminary frequency scan of each radical's Larmor precession frequency, taking into account a small shift between the two radicals in the pair due to dipole-dipole interaction, as written in the second term of Eq.~\eqref{eq:hamiltonian}. Combining this with the proposed protocol for the pair's charge state determination \cite{Liu2017a}, our spin state scheme can allow us to distinguish between the different triplet spin states, as well as giving a threshold below which we can safely establish that the RP is in the singlet state. In Fig.~\ref{fig:experimental}(b) we show the original and modified TEER pulse sequence for detecting the exact spin state of the RP. In Fig.~\ref{fig:experimental}(b) the NV's normalized contrast when employing this pulse sequence while varying the time between microwave pulses, $\tau$, is shown and the four spin states of the two-electron system (RP) become clearly distinguishable. The NV decoherence time and RP spin relaxation time are both in the relevant range for such TEER sensing \cite{Schlipf2017}.

The experimental difficulty in clearly resolving the singlet triplet oscillations stems from the ensemble dynamics of the RP spins in the presence of external field, as shown in  Fig.~\ref{fig:theoretical}(d). One can also use the dynamical control pulses on the sensor spin to resolve its coupling to individual RPs. For this we choose the computational basis for the sensor spin as $\ket{-1}, \ket{0}$. Due to this the effective field seen by a given  RP is $\omega - g$. By scanning the external field $\omega$, one finds an enhanced singlet fraction when the total effective field becomes zero i.e., the singlet fraction $\Phi_S$ peaks at the external field, with an uncertainty (width) determined by the recombination rate $\tilde{\kappa}$. The aspect of quantum control here arises when one periodically changes the spin state of the sensor from $m_s = +1$ to $m_s = -1$ in a controllable way.

\section{Conclusion and Outlook}
The use of a quantum sensor, i.e., a true two-level system as the detection tool opens a wide range of possibilities for optimization and control which are not accessible to macroscopic or classic objects. Namely, it allows us to decouple the target spin from its surrounding spin bath and even make use of potentially helpful nuclear spins in the immediate vicinity of the radicals, such as those which are hyperfine coupled to them.  We find that even such a basic protocol (before optimization) can lead to a visible enhancement of the triplet/singlet ratio, and hence in effect we show control over at least one stage of the radical pair mechanism. As mentioned above, this can be further improved by polarizing the radicals' neighboring nuclear spins.  A fully polarized nuclear spin species, for example could result in nearly a two-fold increase in the sensor contrast when compared to  its thermal polarization. Moreover, due to the quantum nature of both the NV and the RP, it is possible to devise a model which takes such a single spin sensor and an $N$-sized collection of radicals pairs, and show that as one reduces the number of pairs in the sensor's vicinity, one can obtain a significantly improved visibility contrast for reading out the state of the pairs.

To conclude, we have shown here the role of a quantum sensor in sensing and controlling the RP mechanism to weak external fields, both for the case of single and ensemble RPs. We have also given modified multi-electron electron resonance spectroscopy pulse sequences to clearly distinguish the four different singlet/triplet configurations, and find the conditions for the optimal sensitivity in terms of magnetic field strength, hyperfine interaction and recombination rate. We have also shown a preliminary
 quantum control aspect wherein modulation of sensor spin state can have dramatic effect on the RP dynamics and we envisage that additional tailor-made protocols can improve this even further. Our results continue a line of theoretical proposals for detecting various aspects of the phenomenon known as the radical pair mechanism, which can now be considered for experimental realization as the technical aspects of assembling the appropriate setups are being constantly tackled and resolved.

\paragraph*{Acknowledgements}
This research is made possible in part by the historic generosity of the Harold Perlman Family. It is supported by research grants from the Abramson Family Center for Young Scientists, the Ilse Katz Institute for Material Sciences and Magnetic Resonance Research and the Willner Family Leadership Institute for the Weizmann Institute of Science. DD would like to acknowledge the support from DFG (FOR 2724).
\bibliographystyle{nature}
\bibliography{NVSPM}

\appendix
\section*{Appendices}
\section{Estimation for the magnetic field of a radical pair}
The magnetic field from the radical pair is assumed to behave like a magnetic dipole, 
$$
    \mathbf{B} = \frac{\mu_0}{4\pi}\left[\frac{3\mathbf{r}\left(\mathbf{m}\cdot\mathbf{r}\right)}{r^5} - \frac{\mathbf{m}}{r^3}\right],
$$
and so for a best case scenario (where the angle between $\mathbf{m}$ and $\mathbf{r}$ is $\pi/2$), the magnetic field of an electron at a distance of 20 nm is 59 nT.
\color{black}
\section{Dynamics}
The time-evolution operator corresponding to the Hamiltonian (Eq.~\eqref{eq:simp_H}) is given by
\be
U = U_0 \ket{0}\bra{0} + U_{\pm 1} \ket{\pm 1}\bra{\pm 1} .
\ee
where $U_0 = {\rm e}^{iH_0t}$ and $U_{\pm 1} = {\rm e}^{iH_{\pm 1}t}$. Starting from an initial state of the sensor in $\ket{\psi}_S = \frac{1}{\sqrt{2}}[\ket{0}+\ket{1}]$, and the RP in a singlet state  $\ket{\psi}_{\sigma} = \frac{1}{\sqrt{2}}[\ket{\uparrow \downarrow}-\ket{\downarrow \uparrow}]$, the time-evolved state for the total system can be found using 
\be
\rho(t) = U\rho(0)U^\dagger
\ee
where $\rho_{(0)} = \ket{\psi}_S\bra{\psi}_S\otimes\ket{\psi}_{\sigma}\bra{\psi}_{\sigma}\otimes\frac{1}{2}{\hat{I}}_A$,
where the nuclear spin state is taken to be a thermal state, i.e., a fully mixed states $\frac{1}{2}\hat{I}_A$.
The reduced state of RP spins and their coupled nuclear spins can be obtained by tracing out the sensor degrees of freedom, and vice-versa respectively as
\be
\rho_{\sigma, I_A}(t)= \mathrm{Tr}_S\rho(t), ~~ \rho_S(t)= \mathrm{Tr}_{\sigma, I_A}\rho(t).
\ee
For the above initial state of the sensor, the RP state
\be
\rho_{\sigma, I_A}(t)=\frac{1}{2}[U_0\rho_{\sigma, I_A}(t)U_0^\dagger +U_1\rho_{\sigma, I_A}(t)U_1^\dagger ]
\ee
Due to the above symmetry , the total singlet fraction will also be $\Phi_S = (\Phi^1_S + \Phi^0_S)/2$.

\section{Triplet fraction}
Based on the one-proton radical pair model presented in Ref.\,\cite{Timmel1998}, we calculate $\Phi_T(t) = 1-\Phi_S(t)$, where 
\begin{multline*}
    \Phi_S(t) = \frac{3}{8} + \frac{1}{8}\frac{\omega^2}{\Omega^2} + \frac{1}{8}\frac{h^2}{\Omega^2}f(\Omega) \\
    + \frac{1}{8}\left[1-\frac{\omega}{\Omega}\right]f\left(\tfrac{1}{2}h + \tfrac{1}{2}\omega + \tfrac{1}{2}\Omega\right) \\
    + \frac{1}{8}\left[1-\frac{\omega}{\Omega}\right]f\left(\tfrac{1}{2}h - \tfrac{1}{2}\omega - \tfrac{1}{2}\Omega\right) \\
    + \frac{1}{8}\left[1+\frac{\omega}{\Omega}\right]f\left(\tfrac{1}{2}h - \tfrac{1}{2}\omega + \tfrac{1}{2}\Omega\right) \\
    + \frac{1}{8}\left[1+\frac{\omega}{\Omega}\right]f\left(\tfrac{1}{2}h + \tfrac{1}{2}\omega - \tfrac{1}{2}\Omega\right). \\
\end{multline*}

Here $h$ is the hyperfine coupling strength in rad/sec, $\omega$ is the Zeeman interaction term or magnetic field in rad/sec, $\omega = 2\gamma B$, $\Omega = \sqrt{h^2+\omega^2}$ and $f(x) = \cos(xt + \varphi)$. Note that $\varphi$ was not included in the original derivation and accounts for a random phase for each radical pair. We take a magnetic field of 50 $\upmu$T and a hyperfine interaction strength of 14 MHz.

We consider a radical pair initially formed in a singlet configuration, and evolves under both the hyperfine interaction with its neighboring nuclear spins and with the external (Zeeman) field. For the current discussion we neglect their intra-spin interaction and consider their interaction with a nearby probe spin, the NV center. From the coherent evolution of the pair $\rho(t)$ the singlet state dynamics can be evaluated as
\begin{equation}
    P_S(t) = [\ketbra{S_0}{\rho(t)}].
\end{equation}
where $\rho(t)$ is obtained by tracing out the degrees of freedom of the nuclear and probe spins. Further, upon considering the radical-pair recombination mechanisms, the decay of singlet products, or the singlet yield in RP mechanism is simply captured through

\begin{equation}
    \mathcal{P}_S(\kappa) = \int_0^t \mathrm{d}t P_S(t) {\rm e}^{-\kappa t}
\end{equation}

Similarly the contrast as defined in the Eq. (6) can be evaluated to yield an exact expression as given below,

\begin{widetext}
\begin{multline*}
C_\phi({\tilde{\kappa}}) = 2+\frac{1}{\Omega_1 \Omega_2}\left[\tilde{\kappa}  \Omega_1 \left(\frac{4 g \tilde{\kappa}  \Omega_2 (g+\omega )}{2 \Omega_2^2 \left(\tilde{\kappa} ^2-g^2\right)+\left(g^2+\tilde{\kappa} ^2\right)^2+\Omega_2^2}+\frac{2 \tilde{\kappa}  \Omega_2 \left(g^2+\tilde{\kappa} ^2+\Omega_2^2\right)}{\left(\left(g-\Omega_2\right){}^2+\tilde{\kappa} ^2\right) \left(\left(g+\Omega_2\right){}^2+\tilde{\kappa} ^2\right)}\right){h_A^2+\tilde{\kappa} ^2+\omega ^2} \right. \\
+\frac{1}{8} \left(\frac{\tilde{\kappa} }{\left(-g+\Omega_1-\Omega_2\right){}^2+\tilde{\kappa} ^2}+\frac{\tilde{\kappa} }{\left(g+\Omega_1-\Omega_2\right){}^2+\tilde{\kappa} ^2}-\frac{\tilde{\kappa} }{\left(-g+\Omega_1+\Omega_2\right){}^2+\tilde{\kappa} ^2}-\frac{\tilde{\kappa} }{\left(g+\Omega_1+\Omega_2\right){}^2+\tilde{\kappa} ^2}\right) \left(h_A^2+\omega  (g+\omega )\right) \\
\left.-\frac{1}{8} \omega \Omega_2 \left(\frac{\tilde{\kappa} }{\left(-g+\Omega_1-\Omega_2\right){}^2+\tilde{\kappa} ^2}-\frac{\tilde{\kappa} }{\left(g+\Omega_1-\Omega_2\right){}^2+\tilde{\kappa} ^2}+\frac{\tilde{\kappa} }{\left(-g+\Omega_1+\Omega_2\right){}^2+\tilde{\kappa} ^2}-\frac{\tilde{\kappa} }{\left(g+\Omega_1+\Omega_2\right){}^2+\tilde{\kappa} ^2}\right)\right]
\end{multline*}
\end{widetext}

\section{Quantum Control}
The ability to control the singlet product yield through the quantum sensor becomes possible due to the sensor spin-state dependent evolution of the radical pair. Upon coherently flipping of the sensor spin, the singlet yield could be manipulated and hence to its sensitivity to low fields and low recombination rates. To see this, we write down the evolution operator corresponding to the above Hamiltonian, given by
\be
U = U_0 \ket{0}\bra{0} + U_1 \ket{1}\bra{1} .
\ee
Let us consider a $\pi$-flip operation on the sensor spin, followed by the time evolution. This leads to a modified time-evolution operator, given by
\be
V_1 = U{\rm e}^{i\pi S^x}U = U_1U_0 \ket{1}\bra{0} + U_0U_1 \ket{0}\bra{1} .
\ee
Upon the application of $M$ (even) $\pi$-pulses, the final evolution operator takes the simple form
\begin{multline}
V_M = U \cdots {\rm e}^{i\pi S^x}U{\rm e}^{i\pi S^x}U \\ = (U_1U_0)^{M/2} \ket{1}\bra{0} + (U_0U_1)^{M/2} \ket{0}\bra{1} .
\end{multline}
From the above evolution the singlet yield is additionally dependent on the stroboscopic interruptions of evolution for every time $\tau$.

\end{document}